\documentclass[aps,prd,showpacs,nobibnotes,twocolumn,preprintnumbers,
nobalancelastpage,superscriptaddress]{revtex4}
\usepackage{latexsym}
\usepackage{dcolumn}
\usepackage{bm}
\usepackage{amsmath}
\usepackage{natbib}
\input{epsf}
\newcommand{\be}{\begin{equation}}
\newcommand{\ee}{\end{equation}}
\newcommand{\ba}{\begin{eqnarray}}
\newcommand{\ea}{\end{eqnarray}}
\newcommand{\bi}{\begin{itemize}}
\newcommand{\ei}{\end{itemize}}
\newcommand{\bfi}{\begin{figure}[!t]
\epsfxsize=9cm
\epsffile}
\newcommand{\bfig}{\begin{figure*}[t]
\epsfxsize=15cm
\epsffile}
\newcommand{\efi}{\end{figure}}
\newcommand{\efig}{\end{figure*}}
\newcommand{\no}{\nonumber}
\newcommand{\la}{\lesssim}
\newcommand{\ga}{\gtrsim}

\newcommand{\mpch}{{\rm Mpc}/h}
\newcommand{\hmpc}{h/{\rm Mpc}}
\begin{document}
\title{Sampling Artifact in Volume Weighted Velocity Measurement.--- I. Theoretical Modelling}
\author{Pengjie Zhang}
\affiliation{Center for Astronomy and Astrophysics, Department of
  Physics and Astronomy, Shanghai Jiao Tong
  University, 955 Jianchuan road, Shanghai, 200240}
\affiliation{IFSA Collaborative Innovation Center, Shanghai Jiao Tong
University, Shanghai 200240, China}
\email[Email me at: ]{zhangpj@sjtu.edu.cn}
\affiliation{Key Laboratory for Research in Galaxies and Cosmology, Shanghai
  Astronomical Observatory, 80 Nandan Road, Shanghai, 200030,
  China}
\author{Yi Zheng}
\affiliation{Key Laboratory for Research in Galaxies and Cosmology, Shanghai
  Astronomical Observatory, 80 Nandan Road, Shanghai, 200030,
  China}
\author{Yipeng Jing}
\affiliation{Center for Astronomy and Astrophysics, Department of
  Physics and Astronomy, Shanghai Jiao Tong
  University, 955 Jianchuan road, Shanghai, 200240}
\affiliation{IFSA Collaborative Innovation Center, Shanghai Jiao Tong
University, Shanghai 200240, China}


\begin{abstract}
Cosmology based on large scale peculiar velocity preferes volume weighted velocity statistics. However, measuring the volume weighted velocity statistics from inhomogeneously distributed galaxies (simulation particles/halos) suffer from an inevitable and significant sampling artifact. We study this sampling artifact in the velocity power spectrum measured by the nearest-particle (NP) velocity assignment method \cite{paperII}. We derive the analytical expression of leading and higher order terms.  We find that the sampling artifact suppresses the $z=0$ E-mode velocity power spectrum by $\sim 10\%$ at $k=0.1\hmpc$, for samples with number density $10^{-3}(\mpch)^{-3}$. This suppression becomes larger for larger $k$ and for sparser samples. We argue that, this source of systematic errors in peculiar velocity cosmology, albeit severe,  can be self-calibrated in the framework of our theoretical modelling. We also work out the sampling artifact in the density-velocity cross power spectrum measurement. More robust evaluation of related statistics through simulations will be presented in a companion paper \cite{Zheng14a}. We also argue that similar sampling artifact exists in other velocity assignment methods and hence must be carefully corrected to avoid systematic bias in peculiar velocity cosmology. 
\end{abstract}
\maketitle

\section{Introduction}
\label{sec:intro}

Large scale peculiar velocity is maturing as a powerful probe of
cosmology. In particular, peculiar velocity directly responds to gravitational pull of all matter and energy, making it a prominent tool to study the dark universe. Furthermore,  it is sensitive to density inhomogeneities at horizon scales, making
it precious to probe the origin mechanism of the Universe.  

Cosmological applications of peculiar velocity prefer volume weighted velocity statistics. Compared to the density weighted statistics,  the volume weighted one is free of uncertainties in the galaxy density bias.  Unfortunately, the volume weighted velocity statistics is difficult to measure in observations and in numerical simulations.  We can know the velocity where there are galaxies (simulation particles, halos, etc.). But velocity where there are no galaxies (simulation particles) is in general non-vanishing. This sampling artifact inevitably biases the measurement of volume weighted velocity statistics (e.g. \cite{DTFE96,Bernardeau97,DTFE00,Pueblas09,paperII,Koda13}).  It increases  with decreasing particle number density \cite{paperII}. Zheng et al. \cite{paperII} found that it is essentially not a severe problem to measure dark matter (DM) velocity in N-body simulations with particle number density $\bar{n}_P\ga 1 (\mpch)^{-3}$.  But when $\bar{n}_P=0.1 (\mpch)^{-3}$,  it already induces visible suppression at $k\sim 0.2\hmpc$ (Fig. 14, \cite{paperII}). Halos and galaxies in general have lower number density, so the suppression is larger. It increases to $O(10\%)$ at $k=0.1\hmpc$ for  $10^{13}M_\odot$ halos at $z=0$ (\cite{Zheng14a}, hereafter paper II), with $\bar{n}_P\sim 10^{-3}  ({\rm Mpc}/h)^{-3}$.

This sampling artifact can severely bias cosmological constraints. This is obvious for velocity power spectrum measured through sparse galaxy/supernova samples with velocity measurement such as SFI++ and 6dFGS. But it is also the case for redshift space distortion (RSD), which is a major cosmological tool of  Stage IV dark energy projects such as MS-DESI (BigBOSS), Euclid, SKA and WFIRST. These surveys can measure the volume weighted velocity power spectrum through RSD with $O(1\%)$ statistical precision \cite{BigBOSS}. These measurements themselves do not suffer from the above sampling artifact, since the velocity power spectrum is inferred indirectly and statistically from the redshift space galaxy clustering whose measurement is unbiased. However, to constrain cosmology these measurements should be compared to the theoretically predicted velocity power spectrum given a cosmology.  Due to nonlinear evolution of the large scale structure, the most reliable prediction comes from N-body simulations. However, measuring the volume weighted velocity power spectrum in simulations also suffer the same sampling artifact. To match the observational capability of these stage IV projects, velocity power spectrum in N-body simulations must be measured with $O(1\%)$ accuracy at $k\sim 0.1\hmpc$. This is challenging,  requiring through investigation of the sampling artifact. 

This paper presents a theoretical model on the sampling artifact in the volume weighted velocity statistics. Comprehensive study on related statistics and tests against simulations will be presented in paper II. This is along our recent effort to understand peculiar velocity and redshift space distortion \cite{paperI,paperII,paperIII}. When measuring the halo velocity statistics \cite{paperIII}, we find significant sampling artifact. Hence understanding and correcting this sampling artifact becomes prerequisite. Nevertheless, these studies are by themselves technical and are of no cosmological information. Hence we present them in separate papers, instead of in the original series \cite{paperI,paperII,paperIII}.

For brevity, we restrict to the velocity measurement  in N-body simulations, with box size $L_{\rm box}$ and total particle number $N_P=\bar{n}_P L^3_{\rm box}$. The power spectrum is measured on $N_{\rm grid}=N_{\rm side}^3$ grids through FFT.  The sampling artifact is  inevitable no matter what velocity assignment method is adopted, a point that will be elaborated in \S \ref{sec:discussion}. But the details depend on the velocity assignment method.  Throughout this paper, we focus on the sampling artifact in the nearest particle (NP) method \cite{paperII}.  Furthermore, we mainly focus on the volume weighted velocity power spectrum, defined through
$\langle v_i({\bf k}) v_j({\bf k}^{'})\rangle=P_{ij}(k)(2\pi)^3\delta_D({\bf k}+{\bf k}^{'})$.
Here, $i,j=1,2,3$ denote the three components of the velocity field ${\bf v}$. The total velocity power spectrum $P_v(k)\equiv \sum_{i}P_{ii}({\bf k})$.

 The velocity field can be  decomposed into an irrotational and a rotational part. Analogous to the electromagnetic field, we denote the former with subscript ``E'' and the later with subscript ``B''.  In Fourier space, ${\bf v}_E({\bf k})=(\hat{\bf k}\cdot{\bf v})\hat{\bf k}$, ${\bf v}_B({\bf k})={\bf v}-{\bf v}_E$. $P_E\equiv P_{v_E}$ and $P_B\equiv P_{v_B}$ are  the total power spectra of ${\bf v}_E$ and ${\bf v}_B$ respectively, satisfying $P_v=P_E+P_B$. One can verify that $P_E=P_\theta/k^2$ and $P_B=P_w/k^2$, where $\theta\equiv \nabla \cdot{\bf v}$ is the velocity divergence and $w\equiv \nabla\times {\bf v}$ is the velocity vorticity.   We have the following useful relations,
\ba
P_E(k)&=&\sum_{ij} P_{ij}({\bf k}) \hat{k}_i\hat{k}_j\ , \\ 
P_B(k)&=&\sum_{ij} P_{ij}({\bf k})\left(\delta_{ij}-\hat{k}_i\hat{k}_j\right)\ , \no\\
P_{ij}({\bf k})&=&P_E(k)\hat{k}_i\hat{k}_j+\frac{1}{2}P_B(k)(\delta_{ij}-\hat{k}_i\hat{k}_j)\no \ .
\ea
Sampling artifact has very different impact on ${\bf v}_E$ and ${\bf v}_B$. Hence we have to discuss them separately. 

This paper is organized as follows. In \S \ref{sec:model} we list sources of numerical artifacts, including ``shot noise'' due to finite grids (\S \ref{sec:shot}), alias effect and sampling artifact (\S \ref{sec:vps}). In  \S \ref{sec:vps} we derive the analytical expression of the sampling artifact and alias effect, and make clear distinction of the two. Eq. \ref{eqn:PE1} is the most important result of the paper, which describes the leading order sampling artifact in measuring $P_E$. Paper II will show that it is reasonably accurate when $\bar{n}_P\ga 10^{-3}(\mpch)^{-3}$. We also derive higher order corrections, which are required for better accuracy or sparser samples. \S \ref{sec:selfcalibration} provides a recipe to self-calibrate the sampling artifact in $P_E$ measurement, with the aid of our theoretical understanding (e.g. Eq. \ref{eqn:PE1} \& \ref{eqn:PE2}). \S \ref{sec:discussion} argues that no existing velocity assignment methods are  free of sampling artifact, hence understanding and correcting sampling artifact is an essential step in peculiar velocity cosmology. We also derive the sampling artifact in the density-velocity power spectrum measurement in the appendix (\S \ref{sec:appendixB}). 

\bfi{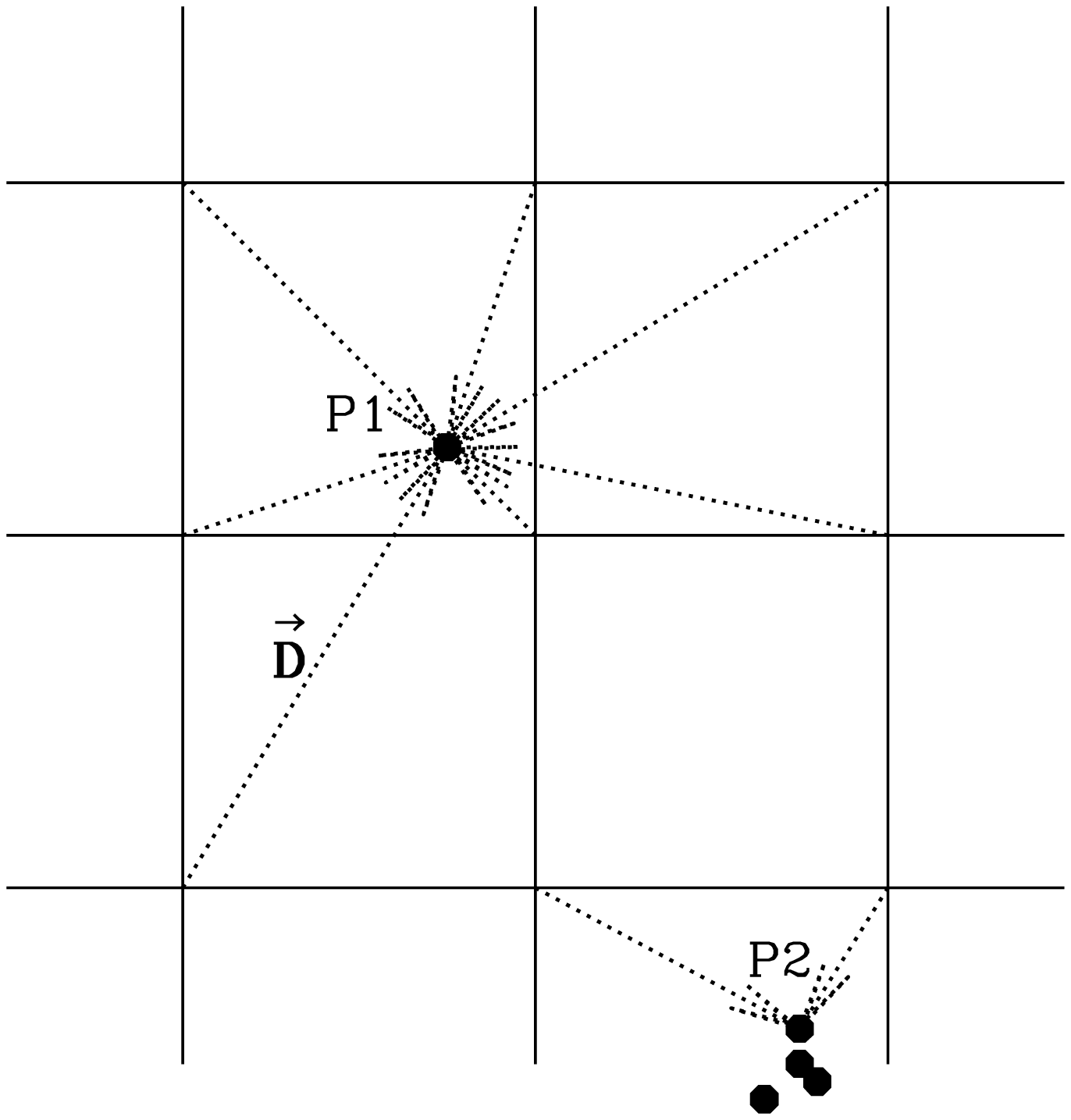}
\caption{The NP method. The points denote simulation particles/galaxies with velocity measurement. The two points with labels ``P1'' and ``P2'' denote two particles who are nearest particles to corresponding grid points.  The separation vector between these particles and corresponding grid points defines  the ``deflection'' field ${\bf D}$. Particles in underdense regions (void), such as ``P1'',  can build correlation between ${\bf D}$ with a correlation length larger than the mean particle separation. Only a fraction of particles in overdense regions are used for the velocity assignment.   \label{fig:NP}}
\efi
\section{Sources of numerical artifacts}
\label{sec:model}
In the NP method (Fig. \ref{fig:NP}), the estimated velocity on the grid position ${\bf x}$ is assigned as that of the nearest particle/halo/galaxy,
\be
\label{eqn:hatv}
\hat{{\bf v}}({\bf x})={\bf v}({\bf x}_P({\bf x}))\ .
\ee
Here, ${\bf x}_P({\bf x})$ is the position of the corresponding nearest particle/halo. 
We recognize the similarity to  CMB lensing \cite{Seljak96}, where lensing alters the photon path, but not the CMB temperature ($T^L(\hat{n}^L)=T(\hat{n})$).
Hence  we can define a ``deflection'' field
\be
{\bf D}({\bf x})\equiv {\bf x}_P({\bf x})-{\bf x}\ .
\ee 
This is the field causing sampling artifact in volume weighted velocity measurement. Previous works such as \cite{Pueblas09,Koda13} discussed the limit of $\bar{n}_P\rightarrow \infty$ (${\bf D}\rightarrow {\bf 0}$). This is the situation for high resolution N-body simulations with $\bar{n}_P\ga 1$ (Mpc/$h$)$^3$. But for halos, the number density is not only much lower, but also a fixed value which can not be increased by increasing simulation resolution. So we have to work on the case ${\bf D}\neq {\bf 0}$ and understand the dependence of sampling artifact on the particle number density. This distinguishes our work from existing ones. 

The statistics we focus on is the velocity power spectrum. As usual we use FFT to obtain the Fourier transform of the velocity field. This is done on discrete uniform grids, 
\ba
\label{eqn:vk}
\hat{{\bf v}}({\bf k})&=&\frac{1}{N_{\rm grid}}\sum_{\bf x} \hat{{\bf v}}({\bf x})\exp(i{\bf k}\cdot {\bf x})
\ea
Here the sum is over all $N_{\rm grid}$ grid points at position ${\bf x}={\bf n} L_{\rm box}/N_{\rm side}\equiv {\bf n} L_{\rm grid}$. Here $L_{\rm grid}\equiv L_{\rm box}/N_{\rm side}$ is the grid size. ${\bf k}={\bf n} 2\pi/L_{\rm box}$, where ${\bf n}=(n_1,n_2,n_3)$ is the 3D integer vector, with $n_i\in (-N_{\rm side}/2,N_{\rm side}/2)$. We then have 
\ba
\label{eqn:vv1}
\langle \hat{{\bf v}}({\bf k}) \hat{{\bf v}}(-{\bf k})\rangle&=&\frac{1}{N^2_{\rm grid}}\sum_{{\bf x},{\bf x}^{'}} \langle \hat{{\bf v}}({\bf x}) \hat{{\bf v}}({\bf x}^{'}) \rangle e^{i{\bf k}\cdot ({\bf x}-{\bf x}^{'})}\no \\
&=&\frac{1}{N^2_{\rm grid}}\sum_{{\bf x}={\bf x}^{'}} \langle \hat{{\bf v}}({\bf x})) \hat{{\bf v}}({\bf x}^{'})\rangle \\
&+&\frac{1}{N^2_{\rm grid}}\sum_{{\bf x}\neq{\bf x}^{'}} \langle \hat{{\bf v}}({\bf x})) {\bf v}({\bf x}^{'}) \rangle e^{i{\bf k}\cdot ({\bf x}-{\bf x}^{'})} \no\ .
\ea

This very ``raw'' measurement of the velocity power spectrum contains three numerical artifacts: shot noise, alias effect and sampling artifact. (1) Unlike the shot noise in the dark matter density field, it does not vanish even if $\bar{n}_P\rightarrow \infty$. (2) The alias effect also arises from the discrete FFT with finite grids. Both the shot noise and the alias effect vanish in the limit $n_{\rm grid}\rightarrow \infty$. (3) In contrast, the sampling artifact arises from the finite $\bar{n}_P$. It vanishes in the limit $\bar{n}_P\rightarrow \infty$, but increases with decreasing $\bar{n}_P$. Hence for DM halos and galaxies with finite $\bar{n}_P$,  this sampling artifact is inevitable and may only be corrected in the post-process. We now proceed to quantify the three effects. 

\section{Shot noise}
\label{sec:shot}
The first term at the right hand side of Eq. \ref{eqn:vv1} is the ``shot noise''. It arises from finite grid points, analogous to shot noise in the dark matter density power spectrum measurement which arises from finite simulation particles.  Unlike the case of dark matter, ``shot noise'' in the velocity power spectrum is part of the signal, the r.m.s of the velocity field. Nevertheless,  it does not arise from spatial correlation in the velocity field, so it should be subtracted to obtain the correct velocity power spectrum. Our first estimation of the velocity power spectrum, with shot noise corrected, is
\be
\label{eqn:shotnoisecorrection}
\hat{P}_{ij}({\bf k})\equiv \langle \hat{v}_i({\bf k})\hat{v}_j(-{\bf k})\rangle V-\frac{P_{\rm shot}(k)\delta_{ij}}{3}\ .
\ee
The normalization $V\equiv L_{\rm box}^3$ is adopted such that the 
velocity dispersion $\int d^3{\bf k}\sum_{ij}\hat{P}_{ij}({\bf k})/(2\pi)^3\rightarrow \sigma_v^2$ in the ideal case ($n_{\rm grid}\rightarrow \infty$ and $L_{\rm box}\rightarrow \infty$). The shot noise power spectrum is, from Eq. \ref{eqn:vv1},
\ba
\label{eqn:shotnoise}
P_{\rm shot}(k)&=&\frac{\sigma^2_{\hat{v}}}{n_{\rm grid}}\ , \\
\sigma^2_{\hat{v}}&\equiv &\frac{1}{N_{\rm grid}}\sum_{\bf x}\hat{v}^2({\bf x})=\frac{1}{N_{\rm grid}}\sum_{\bf x}v^2({\bf x}_P({\bf x}))\no \ .
\ea
Here, $n_{\rm grid}=N_{\rm grid}/L^3_{\rm box}=L^{-3}_{\rm grid}$ is the number density of grid points. 

For coarse grids, this shot noise correction can be significant. Its variance is 
\be
\Delta^2_{{\rm shot}}(k)=5\times 10^3\ ({\rm km/s})^2\frac{\sigma_{\hat{v}}^2}{10^5\ ({\rm km/s})^2}  (kL_{\rm grid})^3\ .
\ee
For the J1200 simulation used in \cite{paperII}, with $N_{\rm grid}=128^3$, $kL_{\rm grid}\simeq 1$ at $k=0.1h/$Mpc and shot noise is already $\sim 10\%$ of the velocity power spectrum. In particular, this shot noise is one of the major contaminations to the ${\bf v}_B$ and ${\bf v}_S$ modes (paper I \& II). 
Shot noise follows the relation
\be
P^{\rm shot}_B(k)=2P^{\rm shot}_E(k)\ .
\ee 
Since both $\sigma^2_{\hat{v}}$ and $n_{\rm grid}$ are directly measurable, shot noise subtraction is straightforward, no need of  prior knowledge on $\sigma^2_{\hat{v}}$. 

We caution that  the naive shot noise subtraction proposed above is imperfect. In particular,  it results in a unphysical constraint,
\be
\label{eqn:sumPk}
\sum_{\bf k}\hat{P}_{ij}({\bf k})=0\ .
\ee
This relation, valid for any $ij$ pair, can be proved combining Eq. \ref{eqn:vv1} \& \ref{eqn:shotnoisecorrection}. (1) It tells us that $\hat{P}_{ii}$ becomes (unrealistically) negative at small scales. This simply means that sparse grids prohibit robust measurement at small scales, so measurement there should not be trusted. To minimize this unphysical behavior, we should choose as large $N_{\rm grid}$ as possible, with the expense of computation speed. Later we will find that,  larger $N_{\rm grid}$ (smaller $L_{\rm grid}$) not only reduces shot noise, but also reduce the alias effect. For this argument, even for sparse halos we would prefer $L_{\rm grid}=5\mpch$ or smaller to measure the velocity at  $k\sim 0.1\hmpc$. (2) It also sheds light on modelling of high order numerical artifacts in $\hat{P}_{ij}$. We will elaborate this point later. 

\section{Shot noise corrected velocity power spectrum}
\label{sec:vps}
The estimated velocity $\hat{{\bf v}}({\bf x})={\bf v}({\bf x}_P({\bf x}))$ is related to the real Fourier component ${\bf v}({\bf q})$ by 
\be
\label{eqn:vq}
\hat{{\bf v}}({\bf x})={\bf v}({\bf x}_P({\bf x}))=\sum_{\bf q} {\bf v}({\bf q})e^{-i{\bf q}\cdot({\bf x}+{\bf D})}\ .
\ee
The wavevector ${\bf q}={\bf n} 2\pi/L_{\rm box}$. Notice that now ${\bf n}$ spanes over the whole integer vector space and that ${\bf k}$ in Eq. \ref{eqn:vk} is the subspace of ${\bf q}$. ${\bf v}({\bf q})$ is the Fourier component that we can measure by FFT with  $N_{\rm grid}\rightarrow \infty$ (but finite $L_{\rm box}$).

Plug Eq. \ref{eqn:vq} into Eq. \ref{eqn:shotnoisecorrection}, we obtain
\ba
\hat{P}_{ij}({\bf k})
&=&\frac{V}{N^2_{\rm grid}}\sum_{{\bf x}\neq{\bf x}^{'},{\bf qq}^{'}}e^{-i({\bf q}-{\bf k})\cdot {\bf x}-i({\bf q}^{'}+{\bf k})\cdot {\bf x}^{'}}\no\\
  &&\times \left\langle  v_i({\bf q})  v_j({\bf q}^{'})e^{-i{\bf q}\cdot {\bf D}-i{\bf q}^{'}\cdot {\bf D}^{'}}\right\rangle\ .
\ea
Here ${\bf D}^{'}\equiv {\bf D}({\bf x}^{'})$. The ensemble average here is not of usual statistics, since it involves both real space (${\bf D}$) and Fourier space properties (${\bf v}({\bf q})$). Nevertheless we adopt the usual condition  ${\bf q}+{\bf q}^{'}=0$ \footnote{For this usual ensemble average, the homogeneity of our universe does not  lead to the condition ${\bf q}+{\bf q}^{'}=0$. It is a complicated issue for future investigation. } and obtain 

\ba
\label{eqn:vv2}
\hat{P}_{ij}({\bf k})&=& \frac{V}{N^2_{\rm grid}} \sum_{\bf q} e^{i({\bf q}-{\bf k})\cdot ({\bf x}^{'}-{\bf x})}\\
&&\times \sum_{{\bf x}\neq{\bf x}^{'}} \left\langle   v_i({\bf q})  v_j(-{\bf q})e^{i{\bf q}\cdot ({\bf D}^{'}-{\bf D})} \right\rangle\no\\
&=&\sum_{\bf q}P_{ij}({\bf q}) W_{ij}({\bf q},{\bf q}^{'})\no\ .
\ea
Here, ${\bf q}^{'}\equiv {\bf q}-{\bf k}$. 
The inhomogeneous window function $W_{ij}$ is defined as
\be
\label{eqn:W}
W_{ij}({\bf q},{\bf q}^{'})\equiv \frac{1}{N^2_{\rm grid}} \sum_{{\bf x}\neq{\bf x}^{'}} S_{ij}({\bf q},{\bf r})_{{\bf D},{\bf v}} e^{i {\bf q}^{'}\cdot {\bf r}}\ .
\ee
Here ${\bf r}\equiv {\bf x}^{'}-{\bf x}$. So $W_{ij}({\bf q},{\bf q}^{'})$ is actually the Fourier transform of the sampling function $S_{ij}({\bf q},{\bf r})_{{\bf D},{\bf v}}$ over ${\bf r}$. 

The sampling function 
\be
\label{eqn:S1}
S_{ij}({\bf q},{\bf r})_{{\bf D},{\bf v}}\equiv \frac{\left\langle  v_i({\bf q})  v_j(-{\bf q})e^{i{\bf q}\cdot ({\bf D}^{'}-{\bf D})} \right\rangle}{\left\langle  v_i({\bf q})  v_j(-{\bf q})\right\rangle}\ .
\ee
$S_{ij}({\bf q},{\bf r}) _{{\bf D},{\bf v}}\neq 1$ means and fully quantifies the sampling artifact, which is caused by ${\bf D}\neq {\bf 0}$. In general $S _{{\bf D},{\bf v}}$ depends both on the ${\bf D}$ field and ${\bf v}$ field. So we explicitly add the subscript ``${{\bf D},{\bf v}}$'' to highlight this  dependence.  

$W_{ij}({\bf q},{\bf q}^{'})$ includes both the alias effect and the sampling artifact. (1) The former is that, since $W_{ij}({\bf q},{\bf q}^{'}\neq {\bf 0})\neq 0$, ${\bf q}\neq{\bf k}$  modes contaminate the power spectrum measurement.  This effect vanishes in the limit $n_{\rm grid}\rightarrow \infty$.  So the alias effect is merely caused by limitation in computation, instead of a fundamental numerical artifact.  (2) The later is that, since the sampling is imperfect (${\bf D}\neq 0$),  $S_{ij}\neq 1$ and $W_{ij}({\bf q},{\bf 0})\neq 1$. It persists in the limit $n_{\rm grid}\rightarrow \infty$. It only vanishes when $\bar{n}_P\rightarrow \infty$ so ${\bf D}\rightarrow {\bf 0}$. But in cases of galaxy or halo distribution, $\bar{n}_P$ is fixed so the above limit  does not apply  in reality and the sampling artifact is inevitable. Hence the sampling artifact is a fundamental numerical artifact. 

The results obtained so far are exact and are applicable to any ``deflection'' ${\bf D}$ field. Fortunately, for realistic ${\bf D}$ field,  Eq. \ref{eqn:vv2} can be significantly simplified. This allows for reasonably accurate estimation of sampling artifact without resorting to numerical calculation through N-body simulations. 

\subsection{The ${\bf D}$ field}
The ${\bf D}$  field depends on the ambient density of particles. So it is a function of both the mean number density of particles $\bar{n}_P$ and local number density fluctuation. The later is a combination of the intrinsic density associated with the large scale structure of the universe,  and Poisson fluctuations due to the discrete particle (halo) distribution. 

\subsubsection{The r.m.s dispersion of the ${\bf D}$ field}
The typical amplitude of ${\bf D}$ is comparable to $L_P\equiv L_{\rm box}N_P^{-1/3}=\bar{n}_P^{-1/3}$, the mean separation between particles. Furthermore, we expect
\be
\sigma^2_{D}\equiv \langle |{\bf D}|^2\rangle=A L^2_P\ .
\ee
The prefactor $A$ in this relation depends on the density fluctuation. Since higher mass resolution (smaller $L_P$) resolve more structures, we expect that $A$ depends on $L_P$. But for $L_P\ga 1h/$Mpc, the Poisson fluctuation is comparable to the intrinsic density fluctuation $\delta$ at such scale. So we only expect a weak dependence on $L_P$. Later we will show that $\sigma_D$ largely determines the leading order sampling artifact. Spatial correlation in ${\bf D}$ leads to higher order effects. 

For a Poisson distribution, we can derive the exact analytical expression of $\sigma_D$. For a given grid point, the probability that there is one particle in the shell $r$-$r+dr$ is $\bar{n}_P dV$, where $V=4\pi r^4/3$. For this particle to be the nearest particle to the given grid point (such that $|{\bf D}|=r$), there must be no particle inside of the sphere of radius $r$. This probability is $(1-p)^{N_P}$. Here, $p=\bar{n}_PV/N_P$ is the probability that a given particle is inside of the sphere. Since $N_P\gg 1$, $(1-p)^{N_P}\xrightarrow {N_P\rightarrow \infty} \exp(-\bar{n}_PV)$. The probability that there is just one particle in the $r$-$r+dr$ shell while no particle locates inside is $\bar{n}_P dV \exp(-\bar{n}_PV)$.  We then obtain
\ba
\sigma_{D}^2(\rm Poisson\  limit)&=&\int_0^\infty r^2\bar{n}_PdV \exp(-\bar{n}_PV)\ .
\ea
The corresponding coefficient $A$ is 
\ba
A(\rm Poisson\  limit)= \left(\frac{3}{4\pi}\right)^{2/3}\Gamma\left(\frac{5}{3}\right)=0.347\ . \no
\ea
Here the Gamma function $\Gamma(t)\equiv \int_0^{\infty}x^{t-1}\exp(-x)dx$.

The real particle distribution is a mixture of Poisson fluctuation and the intrinsic underlying matter density distribution. 
$\sigma_{D}$ increases with the clustering strength of particles. For example, in the limit that all particles locate at a single point, $\sigma_{\bf D}$ is comparable to $L_{\rm box}$. Hence with the presence of intrinsic fluctuation in matter distribution, we have 
\be
\sigma_{D}^2>\sigma_{\rm D}^2({\rm Poisson\  limit})=0.347L_P^2\ .
\ee
Calculating $\sigma_D$ in this case is non-trivial, since in general one can not adopt the Gaussian approximation for the density field. This issue will be further discussed in the appendix \S \ref{sec:appendixA}. 
 Fortunately, there is no need to develop accurate model for $\sigma_D$ because it can be directly measured from simulations/galaxy surveys.

\subsubsection{Spatial correlation in the ${\bf D}$ field}
Exact calculation of $W$ and $S$ requires understanding of spatial correlation in ${\bf D}$, as clearly shown in Eq. \ref{eqn:W} \& \ref{eqn:S1}. Fig. \ref{fig:NP} shows that particles in underdense regions can correlate ${\bf D}$ separated by distance longer than $L_P$, since a particle there can be the nearest particle of several or many surrounding grid points.  ${\bf D}$ can be correlated at even larger scales, due to large scale correlation in $\delta$ \footnote{Another argument is as follows. Let us consider the initial condition in N-body simulation, generated by the Zel'dovich approximation.  If $N_{\rm grid}=N_P$ and if the grid points coincide with the particle position in Lagrangian space, then  ${\bf D}={\bf \Psi}$ where ${\bf \Psi}$ is the Lagrangian displacement. It is curl free, with divergence $\nabla\cdot{\bf \Psi}=-\delta$. This field has a correlation length virtually identical to that of the velocity field, of the order $O(100) $Mpc (e.g. \cite{paperII}). }.

Statistically speaking, $|{\bf D}|$ is smaller in overdense regions and larger in underdense regions (Fig. \ref{fig:NP}). So there is an anti-correlation between the strength of the field $|{\bf D}|$ and the particle number overdensity. Naively we expect 
$\langle |{\bf D}|^2\rangle_V\propto N_V^{-2/3}$. 
Here $\langle \cdots\rangle_V$ denotes averaging over volume $V$. $N_V$ is the total particle number in this volume. This relation fails when $N_V\rightarrow 0$. Nevertheless, it helps understand the ${\bf D}$ field.  $N_V$ has contribution both from Poisson fluctuation and underlying matter density fluctuation $\delta$. In many cases, the two contributions are comparable. 
Both can cause spatial correlation in ${\bf D}$.  It is hard to model theoretically on how large spatial correlation actually is in ${\bf D}$ and how significant  it can affect sampling artifact. However, such theoretical modelling is unnecessary, since relevant statistics can be directly measured through N-body simulations (galaxy data). Numerical evaluation of these statistics will be presented in paper II.

\bfi{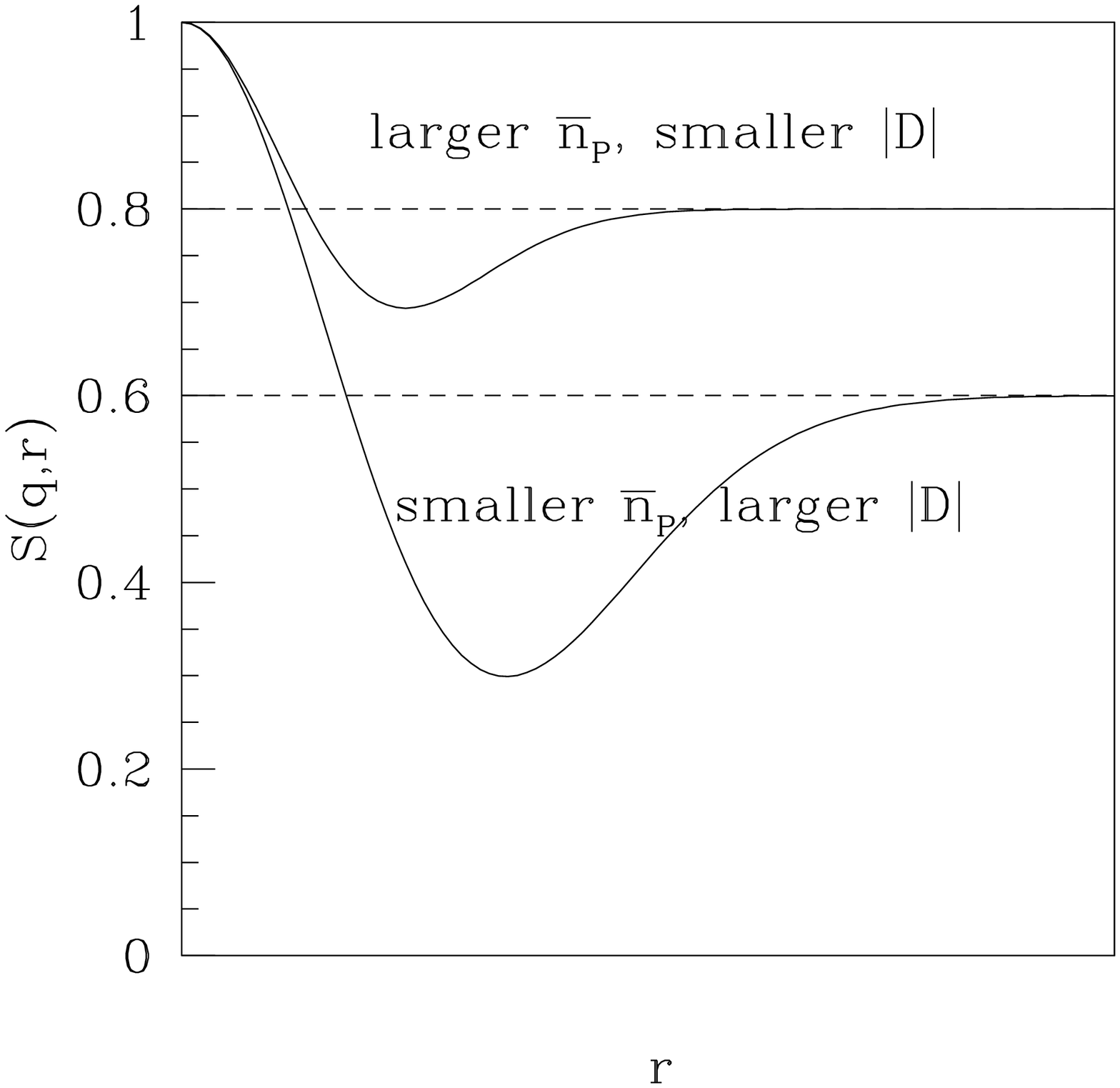}
\caption{A schematic plot of the function $S({\bf q},{\bf r})$ as a function of $r$. It follows that, (1) $S({\bf q},{\bf r}={\bf 0})=1$, (2) $S({\bf q},r\rightarrow \infty)\rightarrow S(q)$ and (3) $\langle S({\bf q},{\bf r})\rangle_{\bf r}=S(q)$. The second and third properties require that $S({\bf q},{\bf r})-S(q)<0$ in intermediate region of $r$.  $S({\bf q},{\bf r})$ decreases with decreasing $\bar{n}_P$, which causes larger $|{\bf D}|$.  \label{fig:S}}
\efi

\subsection{$S({\bf q},{\bf r})$}
In contrast, we only expect weak, if any, correlation between ${\bf v}$ and ${\bf D}$. The same Poisson fluctuation also weakens spatial correlation between ${\bf v}$ and ${\bf D}$. The Fourier component, ${\bf v}({\bf q})$, averages over the whole simulation volume. This further weakens correlation between ${\bf v}({\bf q})$ and ${\bf D}$. Hence hereafter will treat the two as uncorrelated. This significantly simplifies $S_{ij}$ to
\be
\label{eqn:S2}
S_{ij}({\bf q},{\bf r})_{{\bf D},{\bf v}}\rightarrow \left\langle e^{i{\bf q}\cdot ({\bf D}^{'}-{\bf D})} \right\rangle\equiv S({\bf q},{\bf r})\ .
\ee
Now the function $S$ only depends on the field ${\bf D}$ and it is identical for any $ij$ pair. 

$S({\bf q},{\bf r})$ is bounded,
\be
0<S({\bf q},{\bf r})\leq 1\ .
\ee
$S\leq 1$ is trivial to prove and $S=1$  when and only when $r=0$ or $q=0$. $S>0$ can be proved by the cumulant expansion theorem,
\be
S({\bf q},{\bf r}) =\exp\left[\sum_{n\geq 1}\frac{(-1)^n\langle[{\bf q}\cdot({\bf D}^{'}-{\bf D})]^{2n}\rangle_c}{(2n)!}\right]\ .
\ee
Another two useful relations are
\ba
S({\bf q},{\bf r}\rightarrow \infty) &=&\left\langle e^{i{\bf q}\cdot {\bf D}}\right\rangle^2\\
\frac{1}{N^2_{\rm grid}}\sum_{{\bf xx}^{'}}S({\bf q},{\bf r}) \equiv \langle S({\bf q},{\bf r}) \rangle_{\bf r}&=&\left\langle e^{i{\bf q}\cdot {\bf D}}\right\rangle^2\no\ .
\ea
We also define a new function
\be
\label{eqn:Sk}
S(q)\equiv \left\langle e^{i{\bf q}\cdot {\bf D}}\right\rangle^2\ .
\ee
Due to the average, this function only depends on the amplitude, but not the direction of ${\bf q}$. Later we will find that it is a key function describing the sampling artifact. 
By the cumulant expansion theorem, 
\ba
\label{eqn:SA3a}
S(q)&=&\exp\left[\sum_{n\geq 1} \frac{2(-1)^n \langle ({\bf q}\cdot {\bf D})^{2n}\rangle_c}{(2n)!} \right] \no \\
&=&\exp\left[-\frac{1}{3}q^2\sigma_D^2+\cdots\right] \ . 
\ea
The Gaussian approximation includes only the first term at the r.h.s of Eq. \ref{eqn:SA3a}. This approximation is accurate at $q\ll \sigma_D^{-1}\sim L_P^{-1}$. But it beaks at scale $q\ga L_P^{-1}$. In that case,  we recommend the exact definition (Eq. \ref{eqn:Sk}) to evaluate $S(q)$. 

\subsection{$W({\bf q},{\bf q}^{'})$}
$W_{ij}({\bf q},{\bf q}^{'})$ is a derived property of $S_{ij}({\bf q},{\bf r})$. Since now $S_{ij}({\bf q},{\bf r})=S({\bf q},{\bf r})$, $W_{ij}$ is also identical for all $ij$ pairs. Hence we can neglect the subscript and denote $W ({\bf q},{\bf q}^{'})=W_{ij}({\bf q},{\bf q}^{'})$.  Then from Eq. \ref{eqn:vv2} we have  
\ba
\label{eqn:PE}
\hat{P}_E({\bf k})&=&\sum_{\bf q}\left[P_E(q)\cos^2\theta_q+P_B(q)\frac{1-\cos^2\theta_q}{2}\right]\no \\
&&\times W({\bf q},{\bf q}^{'}={\bf q}-{\bf k})\ ,
\ea
\ba
\label{eqn:PB}
\hat{P}_B({\bf k})&=&\sum_{\bf q}\left[P_E(q)\sin^2\theta_q+P_B(q)\frac{1+\cos^2\theta_q}{2}\right]\no \\
&&\times W({\bf q},{\bf q}^{'}={\bf q}-{\bf k})\ .
\ea
$\theta_q$ is the angule between ${\bf q}$ and ${\bf k}$, $\cos\theta_q={\bf q}\cdot{\bf k}/(qk)$.  We further find that 
\ba
\label{eqn:Wps}
W({\bf q},{\bf q}^{'})= \left|\frac{1}{N_{\rm grid}}\sum_{{\bf x}}e^{i{\bf q}^{'}\cdot {\bf x}+i{\bf q}\cdot {\bf D}}\right|^2-N^{-1}_{\rm grid} \ .
\ea
This equation tells us that $W({\bf q},{\bf q}^{'})$ is actually the power spectrum of the field $\exp(i{\bf q}\cdot {\bf D})$. It only requires $O(N_{\rm grid})$ computations to calculate each $W({\bf q},{\bf q}^{'})$ from Eq. \ref{eqn:Wps},  dramatically less than $O(N^2_{\rm grid})$ computations required to calculate  $W({\bf q},{\bf q}^{'})$ from Eq. \ref{eqn:W}. Hence it is Eq. \ref{eqn:Wps} that should be used for numerical calculation of $W$ in simulations. 

However, for analytical discussions, it may be better to use Eq. \ref{eqn:W} instead of Eq. \ref{eqn:Wps}. When ${\bf q}^{'}\neq 2k_N{\bf n}$ where $k_N\equiv \pi/L_{\rm grid}$ is the Neyquist wavernumber, the term $\exp({i {\bf q}^{'}\cdot {\bf r}})$ in Eq. \ref{eqn:W} oscillates around zero when performing the sum over the ${\bf xx}^{'}$ pairs. Since $S({\bf q},{\bf r})>0$, these oscillations largely cancel. In contrast, When ${\bf q}^{'}=2k_N{\bf n}$, $\exp({i {\bf q}^{'}\cdot {\bf r}})=1$, so all $S({\bf q},{\bf r})$ add positively. Hence $W({\bf q},{\bf q}^{'})$ reaches its maximum when ${\bf q}^{'}=2k_N{\bf n}$,
\ba
\label{eqn:Wpeak}
W({\bf q},{\bf q}^{'}=2k_N{\bf n})=W({\bf q},{\bf q}^{'}={\bf 0})\\
=\frac{1}{N^2_{\rm grid}} \sum_{{\bf x}\neq{\bf x}^{'}} S({\bf q},{\bf r})=S(q)-1/N_{\rm grid}\simeq S(q) \ .\no
\ea
We also expect 
\be
\label{eqn:Wpq}
W({\bf q},{\bf q}^{'}\neq 2k_N{\bf n})\ll W({\bf q},{\bf q}^{'}=2k_N{\bf n})\ .
\ee
This inequality can be further quantified by the following relations derived from Eq. \ref{eqn:W}, 
\be
W({\bf q},{\bf q}^{'}+2k_N{\bf n})=W({\bf q},{\bf q}^{'})\ ,
\ee
\be
\label{eqn:sumw}
\sum_{{\bf q}^{'}}W({\bf q},{\bf q}^{'})=0\ ,
\ee
\be
\label{eqn:sumwz}
\sum_{{\bf q}^{'}\in {\bm Z}}W({\bf q},{\bf q}^{'})=0\ .
\ee
This subspace $\bm Z$ is defined with ${\bf q}^{'}=2\pi/L_{\rm box}{\bf m}$ and  $m_{i,{\rm max}}-m_{i,{\rm min}}=N_{\rm side}-1$ ($i=x,y,z$). The above equation means that $W({\bf q},{\bf q}^{'}\neq 2k_N{\bf n})$ oscillates around zero and largely cancels out. Their average is 
\ba
\label{eqn:Wmean}
\frac{\sum_{{\bf q}^{'}\in {\bm Z}, {\bf q}^{'}\neq 2k_N{\bf n}}W({\bf q},{\bf q}^{'})}{N_{\rm grid}-1}&=&-\frac{W({\bf q},{\bf q}^{'}=2k_N{\bf n})}{N_{\rm grid}-1}\\
&\simeq&-\frac{S(q)}{N_{\rm grid}-1} =O(N^{-1}_{\rm grid})\ .\no
\ea
Hence indeed those $W({\bf q},{\bf q}^{'}\neq 2k_N{\bf n})$ are on the average many orders of magnitude smaller than the maximum value $W({\bf q},{\bf q}^{'}=2k_N{\bf n})$.   This relation also tells us that the factor $1/N_{\rm grid}\la 10^{-6}$ in Eq. \ref{eqn:Wps} is non-negligible for evaluating $W({\bf q},{\bf q}^{'}\neq 2k_N{\bf n})$.

Eq. \ref{eqn:Wpeak}, \ref{eqn:Wpq} \& \ref{eqn:Wmean} are the key to derive the leading order alias effect and sampling artifact. 

\bfi{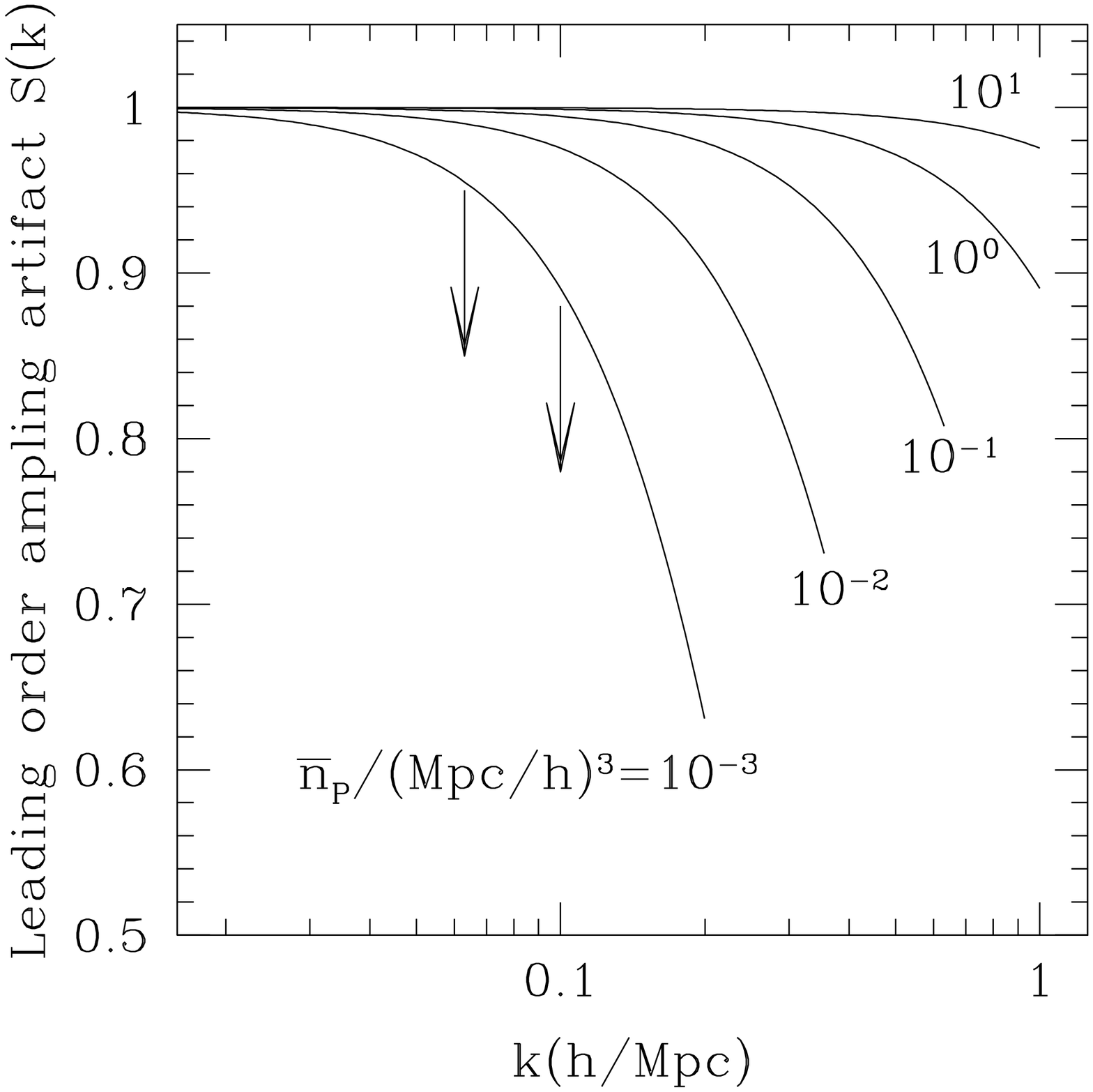}
\caption{The leading order sampling artifact. The measured velocity power spectrum is suppressed by a factor $S(k)<1$.  Both the Poisson fluctuation and intrinsic fluctuation in the particle number distribution cause $S<1$. The figure shown only considers the Poisson fluctuation, so the real sampling artifact is larger than what shown here. Hence this figure shows that sampling artifact in the halo/galaxy velocity power spectrum is significant. It is at the level of $1\%$-$10\%$ at $k=0.1\hmpc$, and at the level of $10\%$ or larger at $k>0.2\hmpc$, when the halo/galaxy number density $\bar{n}_P\la 10^{-2} (\mpch)^3$.  \label{fig:sk}}
\efi

\subsection{Leading order sampling artifact and alias effect}
We are now ready to relate the estimated volume weighted velocity power spectrum $\hat{P}_{ij}({\bf k})$ to the correct velocity power spectrum $P_{ij}({\bf q})$, through summing over all ${\bf q}$s by Eq. \ref{eqn:vv2}.  The most significant contribution comes from ${\bf q}={\bf k}$, for two reasons. First,  $W({\bf q},{\bf q}^{'})$ sharply peaks at ${\bf q}^{'}=2k_N{\bf n}$, and quickly oscillates around zero when ${\bf q}^{'}\neq 2k_N{\bf n}$ with zero mean (Eq. \ref{eqn:sumw} \& Eq. \ref{eqn:sumwz}), so the most significant contributions must come from those ${\bf q}={\bf k}+2k_N{\bf n}$. 
Second, since $S(q)$ decreases quickly with $q$ (Fig. \ref{eqn:Sk}) and since $k\ll 2k_N$ for $k$ of interest, the ${\bf n}={\bf 0}$ mode contributes most among all ${\bf q}={\bf k}+2k_N{\bf n}$ modes. Hence the leading order approximation to Eq. \ref{eqn:vv2} is 
\ba
\label{eqn:1st}
\hat{P}^{(1)}_{ij}({\bf k})= P_{ij}({\bf k}) W({\bf k},{\bf 0})=P_{ij}({\bf k}) S(k)\ .
\ea
The last expression neglects $1/N_{\rm grid}$ with respect to $S(k)$.  This is accurate to the order $10^{-6}$ or better, virtually exact.  Since $W<1$ ($S<1$) due to ${\bf D}\neq {\bf 0}$, the  leading order sampling artifact suppresses the velocity power spectrum.

\subsection{Higher order corrections}
In some cases we may need to include higher order corrections beyond the leading order approximation (Eq. \ref{eqn:1st}). They all come from those ${\bf q}\neq {\bf k}$ modes. Hence they are all be treated as the alias effect in general. However, the alias effect here differs significantly from those discussed in the literature \cite{Jing05,Pueblas09,Koda13}, a manifestation of the sampling function $S$ and the underlying ${\bf D}$ field. The ${\bf q}$s can be written as ${\bf q}={\bf k}+2k_N{\bf n}+2\pi{\bf m}/L_{\rm box}$, with either ${\bf n}\neq 0$ or ${\bf m}\neq {\bf 0}$. Here $m_i\in [-N_{\rm side}/2,N_{\rm side}/2]$. Hence the next to leading order corrections have two sources, those with ${\bf n}\neq {\bf 0}$ and ${\bf m}={\bf 0}$, and those with ${\bf n}={\bf 0}$ and ${\bf m}\neq {\bf 0}$ \footnote{Contribution from ${\bf n}\neq {\bf 0}$ and ${\bf m}\neq {\bf 0}$ modes are subdominant, since the velocity power spectrum decreases rapidly towards small scales.}. We then 
\ba
\label{eqn:2nd}
\hat{P}^{(2)}_{ij}({\bf k})&=&\sum_{{\bf n}\neq {\bf 0}}P_{ij}({\bf k}+2k_N{\bf n})  S(|{\bf k}+2k_N{\bf n}|)\\
&+&\sum_{{\bf m}\neq {\bf 0}} P_{ij}({\bf k}+\frac{2\pi{\bf m}}{L_{\rm box}})W\left({\bf k}+\frac{2\pi{\bf m}}{L_{\rm box}},\frac{2\pi}{L_{\rm box}}{\bf m}\right)\ . \no
\ea
We recognize the first contribution as the conventional alias effect \cite{Jing05,Koda13}, but with an extra factor $S<1$ arising from the sampling artifact. The inclusion of these terms is obvious since $W({\bf q},{\bf q}^{'}\equiv {\bf q}-{\bf k})$ peaks at ${\bf q}={\bf k}+2k_N{\bf n}$.  

The second term is a new type of alias effect, which does not show up in the conventional alias effect \cite{Jing05,Pueblas09,Koda13}. Impact of those ${\bf m}\neq {\bf 0}$ modes is heavily reduced due to the factor $|W|\ll 1$. We have to resort to numerical simulation to evaluate its impact.

Inclusion of the second term in Eq. \ref{eqn:2nd} is necessary \footnote{Inclusion of the second term is also necessary to make the very small scale $P_{ij}({\bf k})$ measurement self-consistent. $\hat{P}^{(1)}_{ii}$ is always positive since $S>0$. However Eq. \ref{eqn:sumPk} requires  $\hat{P}_{ii}$ to be negative at sufficiently small scale. Hence higher order terms must be included to reproduce this behavior. Indeed, $W$ can be negative, making $\hat{P}^{(2)}_{ii}<0$ and $\sum_{\bf k}\hat{P}_{ii}({\bf k})=0$ (Eq. \ref{eqn:sumPk}) possible. Nevertheless, this argument only applies to very small scales, where measurement at such scales is already unreliable due to shot noise subtraction and severe sampling artifact. }. (1)  In reality ${\bf D}$ is spatially  correlated, so $W({\bf q},{\bf q}^{'}\neq 2k_N{\bf n})\neq 0$. Poisson fluctuation in the particle distribution causes a typical correlation length $\sim L_P$. For sparse distribution of halos/galaxies with number density $10^{-6}$-$10^{-3}$ (Mpc/$h)^3$, the correlation length can reach $10$-$100$ Mpc$/h$. Intrinsic density fluctuation can cause correlation at even larger scale. Hence if we want  to measure the velocity statistics  to high accuracy ($\sim 1\%$),   contribution from $W({\bf q},{\bf q}^{'}\neq 2k_N{\bf n})$ may be non-negligible.  (2) The velocity vorticity. At large scale, the real velocity field is curl-free. Hence the measured velocity vorticity $\hat{\bf v}_B({\bf k})$ largely comes from the alias effect of ${\bf v}_E({\bf q}={\bf k}+2k_N{\bf n})$ modes \cite{Pueblas09}.  But in principle it can also come from those ${\bf v}_E({\bf q})$ modes with ${\bf q}$ around ${\bf k}$. Because usually $P_E(k)\gg P_E(|{\bf k}+2k_N{\bf n}|)$ when ${\bf n}\neq 0$ \footnote{Let us consider typical grid size $L_{\rm grid}\leq 10$ Mpc$/h$.  At $k>0.1h/$Mpc, $P_E\propto k^{n_E}$ with $n_E>3$ (Fig. 1, \cite{paperII}).  $P_E(k)/P_E(|{\bf k}+2k_N{\bf n}|)>7^3$, for $k=0.1h/$Mpc. At $k>0.2h/$Mpc, $n_E>4$. $P_E(k)/P_E(|{\bf k}+2k_N{\bf n}|)>4^4$, for $k=0.2h/$Mpc. }, the induced B-mode from ${\bf q}\simeq {\bf k}$ E-mode can be non-negligible. 

\subsection{$P_E(k)$}
The real velocity field is dominated by the E-mode at large scale (e.g. \cite{Bernardeau02,Pueblas09,paperII}). So when consider the measured E-mode, we will take the limit of no intrinsic (real) B-mode. The leading order expression of the measured E-mode power spectrum is then
\ba
\label{eqn:PE1}
\hat{P}^{(1)}_E(k)=P_E(k) S(k)\ .
\ea

Now it is clear that the function $S(k)$ describes the dominant effect of sampling artifact. Since $S(k)<1$ (Fig. \ref{fig:sk}), sampling artifact causes underestimation of the E-mode velocity power spectrum. $1-S(k)$ increases with decreasing $\bar{n}_P$. For sparse samples such as $10^{13} M_\odot$ halos with $\bar{n}_P\sim 10^{-3} (\mpch)^{-3}$ at $z=0$, this underestimation reaches $10\%$ at $k=0.1\hmpc$ (Fig. \ref{fig:sk}). Even for denser samples such as dark matter particles in a $L_{\rm box}\sim 10^3\mpch$ and $N_P\sim 1024^3$ simulation typically used for peculiar velocity study, the underestimation reaches at least $1\%$ at $k=0.3\hmpc$ (Fig. \ref{fig:sk}). Hence this sampling artifact is indeed a severe systematic error in measuring the volume weighted velocity statistics. 

This underestimation of $P_E$ has been observed in \cite{paperII}. For example, Fig. 14 of \cite{paperII} shows that, for the J1200 simulation there, the suppression is visible at $k>0.3$ when the particle number density decreases from $\bar{n}_P=1(\mpch)^{-3}$ to $\bar{n}_P=0.5(\mpch)^{-3}$ to $\bar{n}_P=0.1(\mpch)^{-3}$. But fo the G100 simulation there is no visible suppression at $k<1\hmpc$, with $\bar{n}_P\in (100-10^3) (\mpch)^{-3}$. These behaviours are qualitatively consistent with our expectation (Fig. \ref{fig:sk}). 

Tests against N-body simulation have found good agreement for $\bar{n}_P>10^{-3} (\mpch)^{-3}$ (paper II).  We will leave quantitative comparison in this companion paper. In particular, we will test if Eq. \ref{eqn:PE1} is accurate to the demanded $1\%$ and if higher order corrections are necessary. 

Second order corrections can come from  both the ${\bf n}\neq {\bf 0}$ modes and  the ${\bf m}\neq {\bf 0}$ modes in $\hat{P}_{ij}^{(2)}$ (Eq. \ref{eqn:2nd}). Again taking the limit of no intrinsic (real) velocity B-mode, we obtain
\ba
\hat{P}_E^{(2)}({\bf k})&=&\sum_{{\bf n}\neq {\bf 0}}P_{E}(q_{\bf n})\cos^2\theta_{\bf n} S(q_{\bf n})\no \\ 
&+&\sum_{{\bf m}\neq {\bf 0}}P_E(q_{\bf m}) \cos^2\theta_{\bf m}W\left({\bf q}_{\bf m},\frac{2\pi}{L_{\rm box}}{\bf m}\right)\ .
\ea
Here, ${\bf q}_{\bf n}\equiv {\bf k}+2k_N{\bf n}$ and $\theta_{\bf n}$ is the angle between ${\bf k}$ and ${\bf q}_{\bf n}$. ${\bf q}_{\bf m}\equiv {\bf k}+2\pi/L_{\rm box}{\bf m}$ and $\theta_{\bf m}$ is the angle between ${\bf k}$ and ${\bf q}_{\bf m}$. As a reminder, ${\bf m}$ is bounded with $m_i\in [-N_{\rm side}/2,N_{\rm side}/2]$. 

Since the velocity power spectrum is concentrated at large scale, the alias effect from ${\bf n}\neq {\bf 0}$ modes can be safely neglected, for $L_{\rm grid}\la 5\mpch$. For such grid size, $2k_N> 1.2h/$Mpc. At $k>0.1h/$Mpc, $P_E(k)\propto k^{-n_E}$ with $n_E>3$ (e.g. Fig. 1, paper II \cite{paperII}). So the dominant alias effect comes from the six modes with $|{\bf n}|=1$. At $k=0.1h/$Mpc, the impact is $\la (6\times 12^{-3})S(2k_N)\simeq 0.3\% S(2k_N)$ of $P_E$. At $k>0.2h/$Mpc, $n_E>4$. So the alias effect at $k=0.2h/$Mpc is $\la (6\times 6^{-4}) S(2k_N)\simeq 0.5\% S(2k_N)$. Furthermore we have $S(2k_N)\ll 1$ (Fig. 3) for halos/galaxies.  Since velocity measurement can rarely reach this level of accuracy (but see \cite{McDonald09}), the alias effect from ${\bf n}\neq {\bf 0}$ modes may be safely neglected in any order in realistic applications \footnote{This is in sharp contrast to the alias effect in the density power spectrum, which is significant \cite{Jing05}. This is caused by that the density power spectrum is bluer, with a power index $n_\delta\geq n_E-2$. So the fractional contribution from ${\bf n}\neq 0$ can be orders of magnitude larger.}. Furthermore, if needed, finer grids can be adopted to reduce such alias effect. 

What about the second term?  Although $S(q_{\bf n})$ (${\bf n}\neq {\bf 0}$) can be larger than $W({\bf q}_{\bf m},2\pi {\bf m}/L_{\rm box})$ \footnote{But it is not always correct that $S(q_{\bf n})\gg W({\bf q}_{\bf m},\frac{2\pi}{L_{\rm box}}{\bf m})$. $W({\bf q},{\bf q}^{'})$ decreases sharply with increasing $q$. Since $q_{\bf n}\gg q_{\bf m}$, it is not obvious that $S(q_{\bf n})=W({\bf q}_{\bf n},{\bf q}^{'}={\bf 0})\gg W({\bf q}_{\bf m},{\bf q}^{'}\neq {\bf 0})$.}, the second term can still be larger than the first term, since $P_E(q_{\bf m})\gg P_E(q_{\bf n})$ (${\bf n}\neq {\bf 0}$).  Hence for safety we keep this term and obtain
\ba
\label{eqn:PE2}
\hat{P}_E^{(2)}({\bf k})\simeq \sum_{{\bf m}\neq {\bf 0}}P_E(q_{\bf m}) \cos^2\theta_{\bf m}W\left({\bf q}_{\bf m},\frac{2\pi}{L_{\rm box}}{\bf m}\right)\ .
\ea
Later we will find that these terms can be self-calibrated such that $P_E$ measurement is not biased by these terms. 

\subsection{$P_B(k)$}
The measured B-mode power spectrum has three sources to leading order, the intrinsic one suppressed by a factor $S(k)$, the alias effect from ${\bf n}\neq {\bf 0}$ modes and from ${\bf m}\neq {\bf 0}$ modes. We have
\ba
\label{eqn:SAB}
\hat{P}^{(1)}_B({\bf k})&=&P_B(k)S(k)+\sum_{{\bf n}\neq {\bf 0}}P_E(q_{\bf n})\sin^2\theta_{\bf n}S(q_{\bf n}) \no\\
&+&\sum_{{\bf m}\neq {\bf 0}}P_E(q_{\bf m}) \sin^2\theta_{\bf m}W\left({\bf q}_{\bf m},\frac{2\pi}{L_{\rm box}}{\bf m}\right)
\ea
The intrinsic B-mode can be treated as uncorrelated at large scale since it arises from small/nonlinear scales. Then $P_B\propto k^0$ at large scale (e.g. Fig. 14, \cite{paperII}). The second term at the r.h.s of Eq. \ref{eqn:SAB}  also scales as $k^0$ as long as $k\ll k_N$. This result has been found by  \cite{Pueblas09} in the limit of ${\bf D}={\bf 0}$. Eq. \ref{eqn:SAB} shows that it is also valid when ${\bf D}\neq{\bf 0}$.  But the third one is not expect to scale as $k^0$. 

Depending on simulation specifications and scale of interest, either the real B-mode or the fake one can dominate (e.g. Fig. 14, \cite{paperII}. And also \cite{Pueblas09}). For example,  \cite{paperII} showed that numerical artifacts  at $k=0.1\hmpc$ are negligible  for $\bar{n}_P>100(\mpch)^{-3}$. But when $\bar{n}_P=1(\mpch)^{-3}$, numerical artifacts are already significant. Dark matter halos and galaxies with velocity measurement are sparser than $1(\mpch)^{-3}$,  leading to significant or even overwhelming  numerical artifacts. 

\section{Self-calibrating the sampling artifact}
\label{sec:selfcalibration}
Given the significance of sampling artifact, we shall manage to correct it. Since most cosmological information is encoded in ${\bf v}_E$, we focus on the self-calibration of sampling artifact in $P_E$.  We call it self-calibration in the sense that it does not rely on external information/measurement nor strong priors which could interfere/bias cosmological constraints. Here we  propose a two-step procedure to accurately measure $P_E$. 
\bi
\item Step 1. Measure the shot noise subtracted $\hat{P}_E$ (Eq. \ref{eqn:shotnoisecorrection}) and apply the correction 
\be
\hat{P}_E({\bf k})\rightarrow \frac{\hat{P}_E({\bf k})}{S(k)}\ .
\ee
The field ${\bf D}$ is directly measurable, so $S(k)$ can be measured robustly. The proposed correction method is then straightforward to implement.  This already corrects the majority of the sampling artifact.  In paper II we will quantify its performance and clarify whether it is accurate to the required $1\%$ level. When necessary, one can apply step 2 for further correction. 
\item Step 2. Estimate $\hat{P}_E^{(2)}$  according to Eq. \ref{eqn:PE2}, by approximating $P_E(q_{\bf m})$ there with $\hat{P}_E(k)/S(k)$ measured in step 1. All other terms in Eq. \ref{eqn:PE2} are known, so $\hat{P}_E^{(2)}$ can be measured reasonably well  without prior knowledge of the velocity field.  Subtract this term from $\hat{P}_E$ to correct this second order effect,  and then divide the result by $S(k)$ to correct for the first order effect. Namely, 
\be
\hat{P}_E({\bf k})\rightarrow \frac{\hat{P}_E-\hat{P}_E^{(2)}}{S(k)}\ .
\ee
\ei

\section{Discussions and conclusions}
\label{sec:discussion}
One immediate question is whether similar sampling artifact exists in other velocity assignment methods, such as the ones based on Voronoi and Delaunay tessellations \cite{DTFE96}. As addressed in \cite{paperII}, the NP method is essentially the first step of the Voronoi tessellation method. The two only differ that the Voronoi tessellation method has a second step to smooth the velocity field assigned to the grids. Without this second step of smoothing, the two are identical.    So the Voronoi tessellation method suffers from  essentially the same sampling artifact. 

The Delaunay tessellation method also suffers from the sampling artifact. It linearly interpolates the velocities of simulation particles to obtain the velocity at any point inside of the corresponding tetrahedron whose vertices are particle positions. This linear interpolation guarantees the continuity of the velocity field. However, it does not eliminate sampling artifact, which has significant contribution from second and higher order velocity derivatives. To see this point, Taylor expanding Eq. \ref{eqn:vk},
\ba
\hat{v}_\alpha({\bf k})&=&v_\alpha({\bf k})+\frac{1}{N_{\rm grid}}\sum_{\bf x}\exp(i{\bf k}\cdot {\bf x})\times \\
&&\left[\sum_{\beta=1,2,3}v_{\alpha,\beta}D_\beta+\frac{1}{2}\sum_{\beta\gamma}v_{\alpha,\beta\gamma}D_\beta D_\gamma+\cdots\right]\no\ .
\ea
Here all the properties in the bracket are evaluated at the grid position ${\bf x}$.  This Taylor expansion is analogous to the harmonic approach in CMB lensing \cite{Hu00}. 

 From the above Taylor expansion, one can derive the sampling artifact identical to what derived previously in this paper. For example, in the limit of a random ${\bf D}$ field, one can recover Eq. \ref{eqn:PE1}. However, matching to Eq. \ref{eqn:PE1} requires keeping all derivatives of the velocity field. In particular, even matching Eq. \ref{eqn:PE1} at leading order, namely $S(k)\simeq 1-k^2\sigma_D^2/3$, requires keeping the second order derivatives. The Delaunay tessellation method misses second order derivatives (and higher order ones), hence it can not eliminate a significant fraction of sampling artifact. For this reason, sampling artifact in the Delaunay tessellation method is comparable to that in the NP method. But modelling its sampling artifact may be more difficult than that the NP method, due to more complicated kernel of velocity assignment.

The NP method has an advantage of computationally fast. Together with improved understanding and correction of its sampling artifact, the NP method is suitable to measure the volume weighted velocity statistics. 

\section{Acknowledgement}
This work was supported by the National
Science Foundation of China 
(Grant No. 11025316, 11320101002, 11433001), National
Basic Research Program of China (973 Program 2015CB857001), the
Strategic Priority Research Program "The Emergence of Cosmological
Structures" of the Chinese Academy of Sciences (Grant
No. XDB09000000), and the key laboratory grant from the Office of Science and Technology, Shanghai Municipal Government (No. 11DZ2260700).

\appendix
\section{Complexities in modelling the displacement field}
\label{sec:appendixA}
The Poisson fluctuation can cause that no particles reside in the volume, even if the intrinsic underlying matter density is non-zero. Suppose that  $\bar{\delta}_1$ is the intrinsic overdensity averaged over the volume V,  the probability that there is no particle in this volume given a $\delta_1$ is $\exp[-\bar{n}_PV(1+\bar{\delta}_1)]$. The probability that a particle is inside the shell of volume element $dV$ is $\bar{n}_P(1+\bar{\delta}_2)dV$ where $\bar{\delta}_2$ is the intrinsic overdensity averaged over the shell.  The probability that the nearest particle is at distance $r$, is 
\ba
\label{eqn:pr}
P(r)dr=\bar{n}_PdV\int_{-\infty}^{\infty} (1+\bar{\delta}_2)e^{-x(1+\bar{\delta}_1)}P(\bar{\delta}_1,\bar{\delta}_2)d\bar{\delta}_1d\bar{\delta}_2\no\\
=\bar{n}_PdVe^{-x+\sum_j \frac{(-x)^j\langle\bar{\delta}_1^j\rangle_c}{j!}}\left[1+\sum_{n\geq 1}\frac{(-x)^n\langle \bar{\delta}_1^{n}\bar{\delta}_2 \rangle_c}{n!}\right]\ 
\ea
Here $x\equiv \bar{n}_PV$. 
The last expression adopts the cumulant expansion theorem. The subscript ``$c$'' denotes cumulant.  We recognize the exponential term $\exp(-x+\sum_{j\geq 2} (-x)^j\langle \bar{\delta}_1^j\rangle_c/j!$ as the void probability \cite{White79,Balian89}.  Then $P(r$) is the conditional void probability in that it also requires the existence of a particle in the shell surrounding the void. It is tactually the nearest neighbour distribution, first derived by \cite{White79}. Here with the help of the cumulant expansion, we have redrived this result following a different and independent approach. 

The corresponding coefficient $A$ in the relation $\sigma_D^2\equiv \langle r^2\rangle=AL_P^2$ is
\ba
\label{eqn:A}
A&=&\left(\frac{3}{4\pi}\right)^{2/3}\int_0^\infty dx x^{2/3}\\
&&\times e^{-x+\sum_{j\geq 2} \frac{(-x)^j\langle\bar{\delta}_1^j\rangle_c}{j!}}\left[1+\sum_{n\geq 1}\frac{(-x)^n\langle \bar{\delta}_1^{n}\bar{\delta}_2 \rangle_c}{n!}\right]
\no \ .
\ea
All the correlations in $\langle \cdots \rangle$ are determined by the intrinsic density clustering, so they depend on $r$ and hence on $L_P$ ($r=L_P (3x/4\pi)^{1/3}$). Hence $A$ depends on $L_P$. Thus we show that the intrinsic density fluctuation indeeds affects $\sigma_D$. 

The above results are exact. However they are in general impractical to implement in numerical calculation, due to the non-Gaussianity of the density field. Neglecting the non-Gaussian terms in the above results in general leads to misleading/unrealistic consequences. Assuming Gaussianity, we find that $P(r)\propto \exp(-x+x^2\sigma^2_1/2)(1-x\xi_{12})$. Here, $\sigma_1^2\equiv \langle \bar{\delta}_1^2\rangle$ and $\xi_{12}\equiv \langle \bar{\delta}_1\bar{\delta}_2\rangle$, are all functions of $r$.  When $\bar{n}_P\rightarrow \infty$, $P(r)$ becomes negative where $\xi_{12}$ is positive. This unphysical behavior is caused by the Gaussian assumption in which $1+\bar{\delta}_2$ can become negative. $\xi_{12}>0$ means positive correlation between $\bar{\delta}_1$ and $\bar{\delta}_2$. Hence negative $1+\bar{\delta}_2$ means that  $1+\bar{\delta}_1$ is negative on the average. This leads to negative $(1+\bar{\delta}_2)\exp(-x(1+\bar{\delta}_1))$ with amplitude increasing exponentially with $x\propto \bar{n}_P$. A natural step alleviating this problem is to include the skewness term in Eq. \ref{eqn:A} (the $j=3$ term and $n=2$ term). Now $P(r)\propto \exp(-x+x^2\sigma^2/2-S_3\sigma^4x^2/6)(1-x\xi_{12}+x^2\langle \bar{\delta}^2_1\bar{\delta}_2\rangle_c/2)$. The unphysical behavior of $P(r)<0$ when $x\rightarrow \infty$ disappears. Furthermore, contribution from large $x$ is  suppressed exponentially, because the  reduced skewness $S_3>1/2$ at scales of interest \cite{Bernardeau02}. 

This is just one example on the complexities in calculating statistics of ${\bf D}$. Theoretical calculation of correlation function of ${\bf D}$ is more complicated. Fortunately in reality we do not need such complicated theoretical modelling, since related statistics (e.g. $\sigma_D$ and N-point spatial correlation of the ${\bf D}$ field) can be directly measured since the ${\bf D}$ field is known given the particle distribution. 

\section{Sampling artifact in the density-velocity cross power spectrum}
\label{sec:appendixB}
Similar sampling artifact also exists in the density-velocity cross power spectrum measured through simulations or galaxy surveys with velocity measurement.  The density-velocity cross power spectrum measured through redshift space distortion does not suffer form this sampling artifact. However, we still need to compare it to the one measured through N-body simulations in order to accurately constrain cosmological parameters. Hence, it is necessary to quantify the sampling artifact in this statistics. 

We also consider the measurement in N-body simulations through FFT.  Particles are assigned to corresponding grid point through a homogeneous window function $W_\delta({\bf x},{\bf x}_P)=W_\delta({\bf x}-{\bf x}_P)$ to obtain the density at  grid position ${\bf x}$,
\ba
\hat{\delta}({\bf x})=\frac{1}{\bar{n}_PL^3_{\rm grid}}\sum_P W_\delta({\bf x}-{\bf x}_P)-1\no \ . 
\ea
Here the sum ``P'' is over all particles. Its discrete Fourier transform, after straightforward algebra,  is 
\ba
\hat{\delta}({\bf k})=\sum_{{\bf q}={\bf k}+2k_N{\bf n}} W_\delta({\bf q})\delta({\bf q}) -\delta^{\rm Kronecker}_{\bf k,0}\ .
\ea
Here, $W_\delta({\bf q})$ and $\delta({\bf q})$ are the Fourier transforms of $W_\delta$ and the true overdensity $\delta$ respectively, with infinitesimal grid size but finite box size. $\delta^{\rm Kronecker}_{\bf k,0}$ is the Kronecker delta. 
\ba
\langle \hat{\delta}({\bf k}) \hat{v}_i(-{\bf k}) \rangle &\propto&\sum_{{\bf q}={\bf k}+2k_N{\bf n}}W_\delta({\bf q}) \sum_{\bf x}e^{i({\bf q}-{\bf k})\cdot {\bf x}}\\
&\times &\left\langle \delta({\bf q})v_i(-{\bf q})e^{i{\bf q}\cdot D({\bf x})}\right\rangle \no \\
&\propto&\sum_{{\bf q}={\bf k}+2k_N{\bf n}}W_\delta({\bf q}) \left\langle \delta({\bf q})v_i(-{\bf q})e^{i{\bf q}\cdot D({\bf x})}\right\rangle \no\ .
\ea
The last expression holds since the sum is over discrete grid points ${\bf x}=L_{\rm grid}{\bf n}$. Adopting the approximation that ${\bf D}$ is uncorrelated with $\delta({\bf q})$ and ${\bf v}({\bf q})$, we obtain
\ba
\hat{P}_{\delta v_i}({\bf k})&=& \sum_{{\bf q}={\bf k}+2k_N{\bf n}}P_{\delta v_i}({\bf q}) W_\delta({\bf q})S^{1/2}(q)\ .
\ea
The density- velocity E-mode power spectrum is then
\be
\label{eqn:PdE1}
\hat{P}_{\delta v_E}({\bf k})=\sum_{{\bf q}={\bf k}+2k_N{\bf n}}P_{\delta v_E}(q) \cos\theta_{\bf q}W_\delta({\bf q})S^{1/2}(q)\ .
\ee
The measured power spectrum then suffers from (1) smoothing caused by the density assignment ($W_\delta({\bf q})\leq 1$), (2) the alias effect ($\sum_{{\bf q}\neq {\bf k}} $), and (3) the sampling artifact in the velocity measurement ($S(q)\leq 1$). Comparing to the E-mode velocity power spectrum $P_E$, $P_{\delta v_E}$ has more power at small scales. So the alias effect is larger. However, following the argument for the velocity power spectrum, we expect the alias effect in $P_{\delta v_E}$ to be small at scales of interest. For $L_{\rm grid}= 5\mpch$, its impact is  $\sim 12^{-2}=0.7\%$ at $k=0.1\hmpc$, and $\sim 6^{-3}=0.5\%$ at $k=0.2\hmpc$. The actual impact is even smaller, due to the fact that $|\cos\theta_{\bf q}|<1$, $W_\delta<1$ and $S<1$. Hence we expect the following approximation to be accurate for realistic applications,
\be
\label{eqn:PdE2}
\hat{P}_{\delta v_E}({\bf k})\simeq P_{\delta v_E}(k) W_\delta({\bf k})S^{1/2}(k)\ .
\ee
Since both $W_\delta$ and $S$ are known,  we can simply divide the measured $\hat{P}_{\delta v_E}$ by $W_\delta S^{1/2}$ to correct the sampling artifact and obtain the correct measurement of $P_{\delta v_E}$. Furthermore, if needed, we can also correct the alias effect, adapting the procedure of \cite{Jing05} for the density power spectrum. 

\bibliography{mybib}
\end{document}